# Near-infrared Hong-Ou-Mandel interference on a silicon quantum photonic circuit

Xinan Xu[1], Zhenda Xie[1], Jiangjun Zheng[1], Junlin Liang[1], Tian Zhong[2], Mingbin Yu[3], Serdar Kocaman[1], Guo-Qiang Lo[3], Dim-Lee Kwong[3], Dirk R. Englund[4], Franco N. C. Wong[2], and Chee Wei Wong[1]

[1] *Optical Nanostructures Laboratory, Center for Integrated Science and Engineering, Solid-State Science and Engineering, and Mechanical Engineering, Columbia University, New York, NY 10027*

[2] *Research Laboratory of Electronics, Massachusetts Institute of Technology, 77 Massachusetts Avenue, Cambridge, Massachusetts 02139*

[3] *The Institute of Microelectronics, 11 Science Park Road, Singapore, Singapore 117685*

[4] *Quantum Photonics Laboratory, Columbia University, New York, NY 10027*

Author e-mail address: xx2153@columbia.edu, cww2104@columbia.edu

**Abstract:** Near-infrared Hong-Ou-Mandel quantum interference is observed in silicon nanophotonic directional couplers with raw visibilities on-chip at 90.5%. Spectrally-bright 1557-nm two-photon states are generated in a periodically-poled $KTiOPO_4$ waveguide chip, serving as the entangled photon source and pumped with a self-injection locked laser, for the photon statistical measurements. Efficient four-port coupling in the communications C-band and in the high-index-contrast silicon photonics platform is demonstrated, with matching theoretical predictions of the quantum interference visibility. Constituents for the residual quantum visibility imperfection are examined, supported with theoretical analysis of the sequentially-triggered multipair biphoton contribution and techniques for visibility compensation, towards scalable high-bitrate quantum information processing and communications.

## 1. Introduction

In recent years, quantum information has been popular for its robust applications on cryptography [1–5], computation [6–8] and communication [9,10], and chip-scale cavity quantum electrodynamics [11] involving single photons and single excitons [12–17]. Working with biphoton or multiphoton states and atom-photon interactions, entanglement in various degrees of freedom [17–20], such as time-energy [21,22], spatial-momentum, and polarization [23] has been utilized to harness the efficiency and complexity of quantum information processing. In parallel, quantum secure communications with various protocols [1–3,5,10,24–27], has been proposed to enhance the security of channels and networks. Recent breakthrough experiments are typically achieved in free-space [28], while recent theoretical in-roads on photon transport on-chip [17,29–33] have led in studies on quantum information processing and communication. Emerging measurements of entangled photons on-chip [34–39] have benefited from the arrayed scalability in the nanophotonics platform and potentially robust phase-sensitivity of chip-scale samples albeit with the challenges of device nanofabrication, design, and low-fluence single photon level measurements against chip-scale Rayleigh-scattering photon and coupling losses. In the silica system with remarkable phase control, visibilities up to 98.2% were observed [34]; in the compact silicon system, raw visibilities up to 80% were observed [35]. Most chip-scale measurements have been performed at the visible wavelengths and with bulk nonlinear crystal sources, although there are some recent instances at near-infrared and telecommunications wavelengths [40–42].

Here we report observations of near-infrared Hong-Ou-Mandel (HOM) quantum interference in chip-scale silicon nanophotonics circuits, introducing the biphoton experiments to the integrated optics regime. Employing spectrally-bright type-II periodically-poled KTiOPO$_4$ waveguides (PPKTP) as the entangled photon source, we demonstrate raw quantum visibilities up to 90.5% on-chip – one of the highest visibilities observed in the silicon CMOS-compatible platform. Furthermore, we evaluate the various sources of residual distinguishability including multiphoton pairs, chip-scale excess loss and non-ideal splitting ratios, and polarization effects. The observed interference visibility matches our theoretical predictions, for the different symmetric and asymmetric integrated directional couplers examined.

**2. Near-infrared Hong-Ou-Mandel experimental setup**

Figure 1 illustrates the experimental setup. A 1-cm periodically-poled KTiOPO4 waveguide [43] from AdvR serves as the source for indistinguishable photons [44]; in this case, the waveguide is poled and designed for quasi-phase-matching and high-fluence spontaneous parametric downconversion (SPDC) at approximately 1556-nm to 1558-nm wavelengths. We use a relatively high power (100-mW; QPhotonics QLD-780-80S) semiconductor laser diode as the pump for sufficiently high biphoton rates at approximately $10^7$ per second, to compensate for losses in the fiber and free-space chip coupling setup. The laser is thermally-tuned and stabilized by self-injection locking to 778.9-nm, which is exactly half of the center working wavelength of the PPKTP waveguide. The temperature of the PPKTP waveguide is typically controlled to ~ 25°C for optimal phase matching. A long-pass-filter with cutoff at 1064-nm (Semrock BLP01-1064R-25) blocks pump photons after the SPDC process, and a band-pass filter with 3-nm (Semrock NIR01-1570/3-25) passes the non-degenerate biphoton states. The polarization controller right before the fiber-based PBS is used to tune the polarization so that the fiber-based polarization beamsplitter (PBS) spatially separates the correlated photons. In one branch, a tunable delay is realized by a retroreflector (Thorlabs PS971-C) and a picomotor stage with loss less than 1-dB. In both branches, polarization controllers are introduced to respectively change the polarization of each channel to match the transverse magnetic (TM) mode for coupling into the chip waveguides (Figure 1b).

The chip coupling setup is built with six aspheric lenses, each mounted on individual three-axis precision stages. The two input and output beams are separated by a D-shaped mirror after 60 cm divergence to avoid crosstalk. Single and coincidence measurements are performed by two InGaAs single photon Geiger-mode avalanches detectors $D_1$ and $D_2$ from Princeton Lightwave, with ~ 300 ps gate widths and ~ 20% detection efficiencies. The clock of $D_1$ is set to 15 MHz, and its output signal triggers $D_2$. This allows the coincidence rate to be read directly from the $D_2$ counting rate, with the optical delay calibrated to compensate the electronic delay.

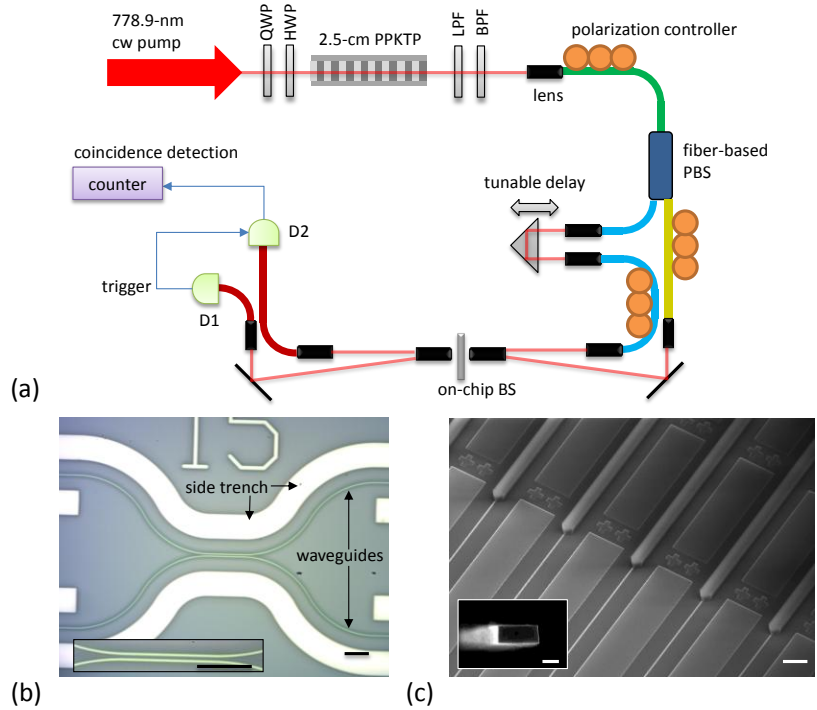

Fig. 1. (a) Experiment setup for near-infrared Hong-Ou-Mandel interference in silicon quantum photonic chip. The pump laser source is realized by an injection-locked semiconductor laser. Fiber polarization controllers are used to ensure biphoton splitting via fiber polarization beam splitter, and to equalize the TM polarization coupling onto the silicon chip. The photon statistics are collected with one single photon detector triggering the other to diminish the dark counts and accidentals. QWP: quarter-wave plate; HWP: half-wave plate; PPKTP: periodically-poled $KTiOPO_4$ waveguide; LPF: low-pass filter; BPF: band-pass filter; PBS: polarization beam splitter; BS: beam splitter. (b) Optical micrograph of nanofabricated directional coupler in silicon-on-insulator. The side trenches (in white) are intended to mark and locate the device. Inset: zoom-in optical micrograph of the waveguide directional. Both scale bars: 1-um. (c) SEM of silicon inverse taper couplers with top oxide cladding waveguides. Scale bar: 20-um. Inset: end-view of protruded silicon waveguide. Scale bar: 200-nm.

### 3. Design and fabrication of silicon chip-scale two-photon interference directional coupler

To ensure good quantum interference on-chip, we examined the design space of the directional couplers, in both transverse electric (TE) and TM polarization states as shown in Figure 2. Differential gap widths ($g$), cross-over coupling lengths ($l_c$) and waveguide widths ($w$) are illustrated for the optimal coupling length and splitting ratios. The silicon waveguides are designed with a 250-nm thickness and for operation at 1550-nm wavelengths.

To calculate the phase velocity of different polarization and symmetry, we use the frequency-domain Maxwell equation fully-vectorial eigenfrequency solver (MPB), which computes by preconditioned conjugate-gradient minimization of the block Rayleigh quotient in a planewave basis [45]. The cross-over coupling length $l_c$ of the two waveguides is then represented as $l_c = \pi/(v_{p,sym} - v_{p,anti,sym})$, in which the phase change of $\pi$ between the symmetric mode and anti-symmetric mode [46] allows for complete crossover from one waveguide to another [47] in an ideal scenario. For a perfect 50-50 splitting ratio, the desired length for the coupler should be

$$L_{c,3dB} = \frac{(2n-1)\pi}{2(v_{p,sym} - v_{p,anti-sym})} + l_{eff}, n = 1, 2, 3... \quad (1)$$

in which $l_{eff}$ is the effective coupler length for the incoming and outgoing bend regions, which can be estimated by an integral of coupling length as a function of gap size along the bending region and computed to be 3-um in our designs (Figure 1b). In addition to the MPB and integral computations, the designs were examined with both rigorous finite-difference time-domain computations and semi-vectorial BeamPROP method from RSoft. With the birefringent character of the directional coupler, we work with the TM mode rather than the TE mode due to its shorter coupling length and greater length control sensitivity. Furthermore, our simulation models and experimental measurements confirm lower loss in the TM mode for straight waveguide as well as the directional coupler regime due to lower electromagnetic field amplitude at the sidewalls (typically rougher than the top and bottom surfaces) [48–51]. The lower loss helps to increase the coincidences count rates and reduce the internal phase shift fluctuations of directional coupler. A quantitative calculation suggests the loss of TE mode is 7.4 times higher than TM mode for a consistent sidewall roughness. In one optimized instance, the waveguide width and coupler length for TM symmetric splitting is chosen to be 400-nm and 15-um, respectively, as illustrated in Figure 2 (Design 1). In this design, the corresponding TE-polarization splitting ratio imbalance was numerically computed to be 9-dB. The excess loss at the optimized directional coupler of Design 1 is estimated to be 0.1-dB by finite-difference time-domain computations.

Further increasing the coupler length will change the splitting ratio imbalance (SR), which could be determined by:

$$SR = \left| \frac{1 + e^{\frac{i\pi(l_{eff}+l)}{l_{couple}}}}{1 - e^{\frac{i\pi(l_{eff}+l)}{l_{couple}}}} \right|^2 \quad (2)$$

For a general comparison, we illustrate and select two other directional couplers with 28-um and 30-um coupling lengths for experimental comparison (Figure 2, Designs 2 and 3). These designs have splitting ratio imbalances corresponding to 2.3-dB and 7.7-dB respectively. Such a large splitting ratio imbalance will remove the indistinguishability and enable the path information, potentially modifying the Hong-Ou-Mandel dip visibility, given by $V = 2SR/(1+SR^2)$ [34,52]. The visibility is 100% for a perfect beamsplitter but is estimated to reduce to 97%, 80% and 47% for splitting ratio imbalances of 1-dB (1.27×), 3-dB (2×), and 6-dB (4×) respectively. For balanced chip-scale splitting, we note that multi-mode interference [35] and Y-splitters are also good elements for physical realization. Directional couplers on the other hand provides differential and accurate thermal tuning on the SR, enabling controlled asymmetries such as for various C-NOT gate [53], quantum cloning [54,55], and Fock state filtration [56,57] applications.

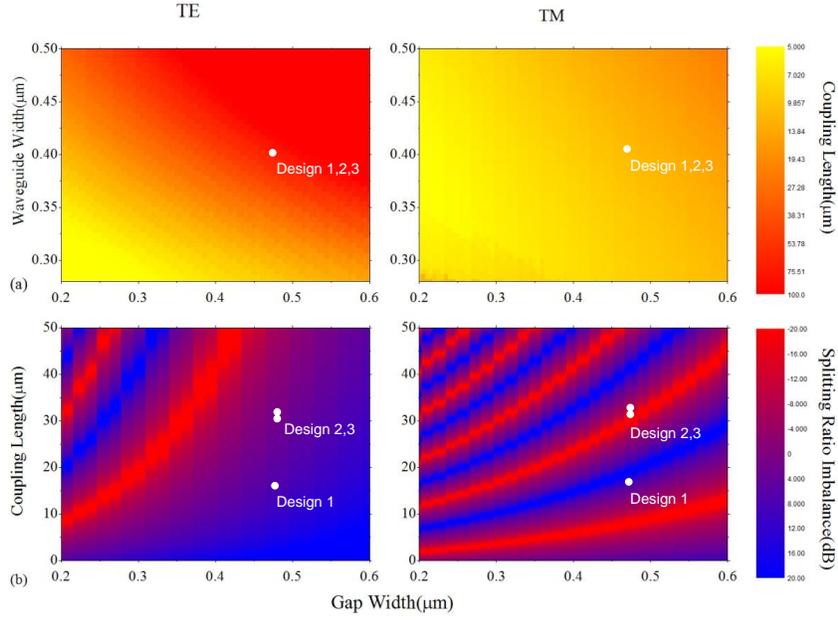

Fig. 2. Design map of silicon photonic directional coupler for two-photon interaction, in both transverse electric (TE; left panels) and transverse magnetic (TM; right panels) polarizations. Panel (a): cross-over coupling length ($l_c$) versus directional coupler gap widths (g) and waveguide width (w). Panel (b): splitting ratio versus designed cross-over coupling length $l_c$ and g. The device thickness is fixed at 250-nm on a thick (typically 3-um) silicon oxide, and the biphoton state input center wavelength is in the 1550-nm telecommunications band. The discretization in each of the panels is from finite numerical simulations. The white circle points denote the designed and fabricated device choices.

Supported by these designs, the devices were next fabricated at the Institute of Microelectronics. Silicon-on-insulator wafers were used, with 248-nm deep-ultraviolet lithography for resist patterning. Sidewall roughness was minimized by optimized lithography, resist development and etching. The measured linear scattering loss of 3-dB/cm in the channel waveguides is determined by folded-back (paperclip) waveguide structures. The inverse couplers are implemented with a tapered silicon nanotaper [58] and top oxide cladding as shown in Figure 1c. The samples are diced and prepared for measurement. The typical total lens-chip-lens coupling loss is approximately 11-dB, or a -14-dB transmission including the -3-dB on-chip splitting. With a measured waveguide propagation loss of 3-dB/cm, the estimated facet coupling loss is 4-dB/facet. Taking into account the waveguide-to-fiber coupling, transmission efficiencies of optical components, and detector efficiencies, the overall single photon detection efficiency is estimated near 1%.

**4. 1557.8-nm Hong-Ou-Mandel visibilities on-chip**

For a pump power of 2.5-mW, the single photon rates coupled in the four-port chip are determined to be about 1000 per second, with dark count rates around 200 per second. The coincidence rate is about 1 pair per second through the silicon photonic chip, with about 1/600 accidental photon pairs per second. With our sequential triggering approach (detector $D_2$ triggered by $D_1$), instead of time-tagging, the coincident dark counts are negligible. An example coincidence versus the relative optical delay is illustrated in Figure 3a, with the observed near-infrared Hong-Ou-Mandel quantum interference on-chip.

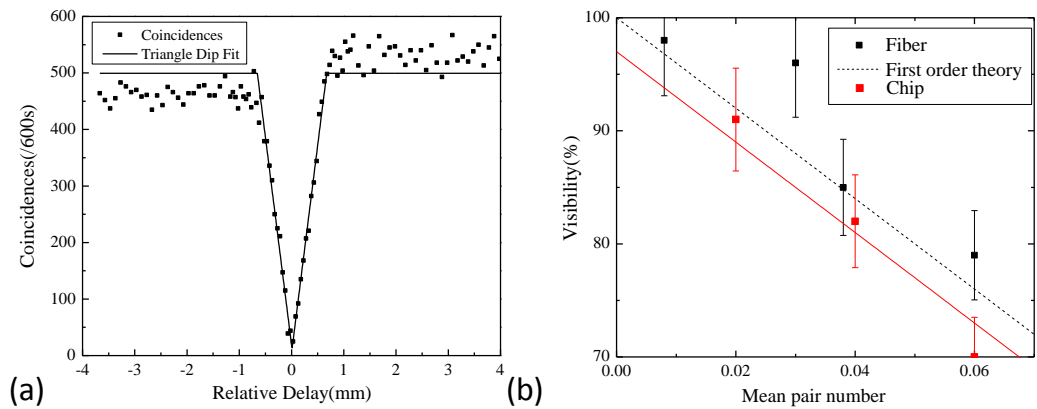

Fig. 3. (a) Coincidences measured on the optimal directional coupler chosen experimentally with a splitting ratio (SR) less than 1-dB. A triangle fit is used for visibility estimation. A raw visibility of 90.5% is observed without accidental subtraction, and 90.8% with accidentals subtraction. (b) Visibility measured with different pump powers for both chip and fiber beam splitter implementations, for comparison. The visibility is approximately linearly related to the pump power as more probability of multiple biphoton pairs generated in one gate window. The first order theory is plotted as dashed line. The On-chip visibility is slightly lower than off-chip one by about 3%, which could be considered to be induced by the chip.

These measurements are performed on a device carefully selected from an array of devices, particularly one with splitting ratio imbalance of less than 1-dB. The sweep resolution and integral time near the dip are set at 50-um and 1200-seconds respectively, which are twice higher resolution and integral time compared to that away from the zero-delay point. The resulting long 21-hour measurement results in small coupling drifts with slightly lower coincidence rate on the negative relative delays. The optimized lowest coincidence is 25 per 600 seconds with a swing coincidence (away from the zero-delay point) of 499 per 600 seconds, giving a raw quantum visibility of 90.5%. The visibility is 90.8% after background accidentals subtraction. An inverse triangle fit is used to estimate the shape of the dip. The measured base-to-base width of Hong-Ou-Mandel dip is 1.36 mm ± 0.07 mm, corresponding to two-photon coherence time of 4.53 ps, or an obtained SPDC bandwidth of 1.79 nm.

## 5. Sources of chip-scale interaction distinguishability

To further uncover the sources of distinguishability, we compare the on-chip Hong-Ou-Mandel visibility with that of a fiber beam splitter (without chip) as illustrated Figure 3b. We plot the visibility against different pump powers or the mean photon pair number to estimate the effects of the chip on the visibility. Since a higher pump power with more biphoton pairs will cause a higher probability of multiple biphoton pairs in one detector gate window, the visibility is inversely proportional to the pump power [43].

Here we note that the effect of multipair biphoton generation in our sequential triggering approach is different from the time-tagging approach. For a baseline model, we assume that the two detectors have uniform detection efficiencies, gate widths and response times, with an infinitesimal timing jitter compared to the gate width. Then the probability of $n$ photon pairs generated in the gate time $\tau$ obeys Poisson distribution: $p(t,n)|_{t=\tau}=(\lambda\tau)^n e^{-\lambda\tau}/n!=\alpha^n e^{-\alpha}/n!$, where $\alpha$ is mean pair number within the gate [59]. To maximize the coincidences, the photon transmitted to the triggered detector is delayed by half of the gate time ($\tau/2$) to guarantee it will always appear within the gate whenever the other photon arrives first (Figure 4a). To calculate the swing coincidences, or the probability of the coincidence event when two photons are relatively delayed and totally distinguishable, we consider only one photon pair per gate to neglect higher order terms (Figure 4a):

$$C_{max} = p(\tau,1) \cdot \frac{1}{2} \cdot \eta^2 = \frac{1}{2}\alpha\eta^2 \qquad (3)$$

where the one half denotes the 50% probability that the biphotons will separate to two gates, and $\eta$ denotes the overall detection efficiency, including all losses and intrinsic detector efficiency. To calculate the probability of coincidence when two photons are indistinguishable, we consider only one and two photon pairs within the gate. Here we notice that even when there is only one photon pair within the detection gate of triggering detector $D_1$, there are still some coincidences contributions (Figure 4b):

$$C_{min}(1) = p(\tau,1) \cdot \frac{1}{2} \cdot [1-(1-\eta)^2]^2 \cdot \left\{ \int_0^{\frac{1}{2}\tau} dt \cdot \frac{1}{\tau} \cdot p(\frac{1}{2}\tau-t,1) \cdot \frac{1}{2} + \int_{\frac{1}{2}\tau}^{\tau} dt \cdot \frac{1}{\tau} \cdot p(t-\frac{1}{2}\tau,1) \cdot \frac{1}{2} \right\} = \frac{1}{4}\alpha^2\eta^2(1-\eta) \qquad (4)$$

where the photon pair is considered uniformly distributed within the gate window, and the possible photon pair within the leak window due to gate time mismatch is considered (Figure 4b). If there are two photon pairs within the gate window of $D_1$, there are four possible situations: (a) the first photon pair is in the path to $D_1$, and second photon pair is in the path to $D_2$ (Figure 4c); (b) the first photon pair is to $D_2$, and the second photon pair is to $D_1$; (c) both photon pairs are to $D_2$; (d) both photon pairs are to $D_1$. Thus we have:

$$C_{min}(2a) = p(\tau,2) \cdot \frac{1}{4} \cdot [1-(1-\eta)^2]^2 \cdot \left( \int_0^{\frac{1}{2}\tau} dt \cdot p_1(t) \cdot \left[ \frac{\frac{1}{2}\tau}{\tau-t} + p(\frac{1}{2}\tau-t,1) \cdot \frac{1}{2} \right] + \int_{\frac{1}{2}\tau}^{\tau} dt \cdot p_1(t) \cdot \left[ 1 + p(t-\frac{1}{2}\tau,1) \cdot \frac{1}{2} \right] \right)$$

(5)

$$C_{min}(2b) = C_{min}(2a) \qquad (6)$$

$$C_{min}(2c) = 0 \qquad (7)$$

$$C_{min}(2d) = p(\tau,2) \cdot \frac{1}{4}[1-(1-\eta)^2]^2 \cdot [1+(1-\eta)^2] \cdot \left[ \int_0^{\frac{1}{2}\tau} dt \cdot p_1(t) \cdot p(\frac{1}{2}\tau-t,1) \cdot \frac{1}{2} + \int_{\frac{1}{2}\tau}^{\tau} dt \cdot p_1(t) \cdot p(t-\frac{1}{2}\tau,1) \cdot \frac{1}{2} \right]$$

(8)

Taking the first order approximation, we have that:

$$C_{min}(2) = C_{min}(2a) + C_{min}(2b) + C_{min}(2c) + C_{min}(2d) = \frac{3}{4}\alpha^2\eta^2(1-\eta) \qquad (9)$$

Here, $p_1(t) = (2\tau-2t)/\tau^2$, which denotes the probability distribution of the first arriving photon pair. We notice here the difference between the sequential triggering approach versus the time-tagging approach is that there is a situation that the second photon pair will be located within the gate window of one detector, but is cut off by the gate window of the other detector (Figure 4c). This portion equals to $2p(\tau,2) \cdot 1/4 \cdot [1-(1-\eta)^2]^2 \cdot \int_0^{\frac{1}{2}\tau} dt \cdot p_1(t) \cdot (1/2\tau - t)/(\tau - t)$, which is exactly the same as the contribution of coincidence conditioning only one photon pair per gate (Equation 4) even when disregarding the detection efficiency distribution within the gate and timing jitter. As these two terms compensates each other, we conclude that, to first order, the visibility for the sequential triggering scenario is as same as time-tagging scenario: $V = 1 - 4\alpha(1-\eta)$.

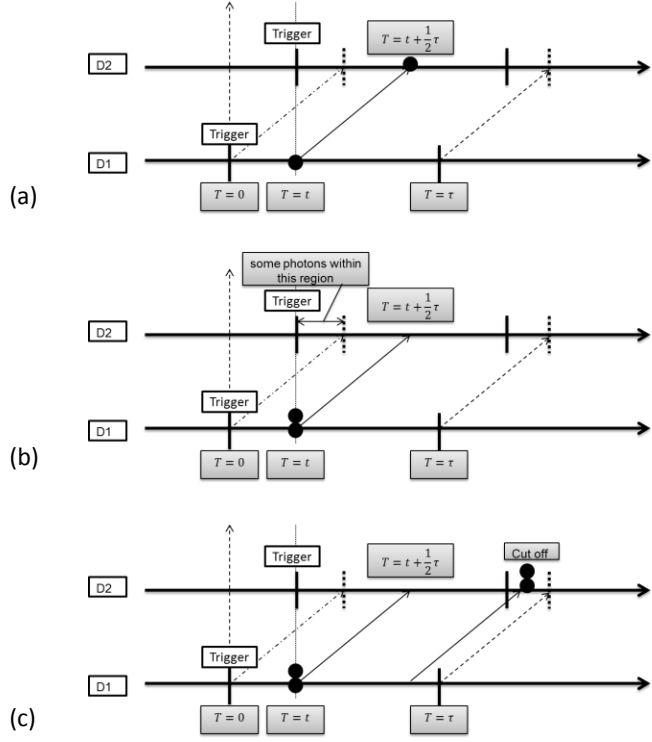

Fig. 4. Scenario of the timeline for the photon pairs. (a) The delay of two photon pairs is set to τ/2 to maximize the coincidences. (b) When there is only one photon pair in the gate window of D1, there is still possibility that D2 will record a photon event due to gate window time mismatch. (c) When there are two photon pairs within the gate window and separated to two detectors, there is possibility that the latter photon pair will be cut off due to the gate window time mismatch.

From fitting the chip result with the same slope as suggested by the above theory, we conclude that 6% of the imperfect visibility is therefore likely to be from the multiphoton pairs. The residual 3% is likely to be induced by processes on-chip. To further understand the chip mechanisms for visibility reduction, we next compared the visibility for different splitting ratios. We selected two devices with coupler lengths of 28-and 30-um, which has the TM mode splitting ratio imbalance of about 3-dB and 6-dB as measured. The comparison of the coincidence measurements between the three silicon chip devices is shown in Figure 5a (before normalization, with lower integral time of 120 seconds compared to Figure 3a). The inverse triangular fit is utilized to estimate the visibility and corresponding deviations. For the 28-um directional coupler, the visibility is measured to be 74 ± 8%, close to the theoretical estimate of 80%. For 30-um directional coupler, the visibility after fitting is 31 ± 11%, compared to the theoretical estimate of 47%, in similar ballpark. The deviations here from theory are due to on-chip directional coupler internal loss and high pump power. For our optimal 15-um directional coupler, the less than 1-dB splitting ratio imbalance (limited by precision of lens-chip coupling loss variations) with its 97% theoretical visibility can therefore account of a sizable portion of the residual 3% decrease in visibility.

Moreover, to understand the quantum interference effect with variation of polarization, we rotate the polarization for one branch of the input path before the chip using a half waveplate. The resulting visibility versus the linear polarization angle is depicted in Figure 5b. The result shows cosinusoidal behavior that reaches maximum visibility with no polarization rotation, and diminished visibility with orthogonal polarization. The maximum visibility in this set of measurements is 83% due to higher pump power of 5-mW. Here we note that the different splitting ratio of TE mode does not affect the visibility, since the probability amplitude of both reflected photons $A_{rr}$ on the biphoton states basis $[|TMTM\rangle, |TETM\rangle, |TMTE\rangle, |TETE\rangle]$ is $\left[\frac{\sqrt{2}}{2}\cos\theta, r_{te}\sin\theta, 0, 0\right]$, and both transmitted photons $A_{tt}$ is $\left[1/\sqrt{2}\cos\theta, 0, t_{te}\sin\theta, 0\right]$. The visibility $1-(|A_{rr}+A_{tt}|^2)/(A_{rr}^2+A_{tt}^2)$ equals to $\cos^2 x$, which does not require the splitting ratio of TE mode as long as TM mode has balanced splitting. In our measurements, the input polarizations are optimized and hence unlikely to be cause of the residual 3% decrease in visibility.

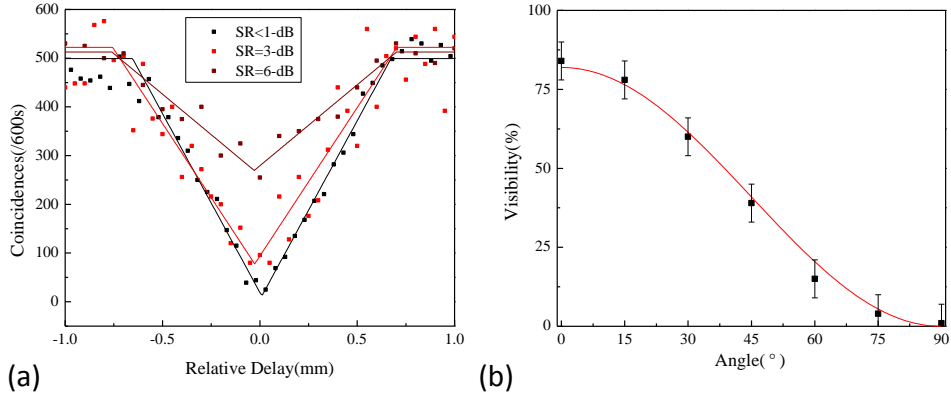

Fig. 5. (a) Coincidences measured on three different directional couplers measured with different splitting ratio imbalances: 6-dB, 3-dB, and less than 1-dB. (b) Visibility versus polarization detuned at one of the input paths.

Another major possible contribution to the chip-induced visibility reduction can be from excess loss of the directional coupler. An ideal free space beamsplitter gives a 180° phase shift for one path of reflection and 0° for the other path, while fiber-based beamsplitter or directional coupler should give both 90° phase shifts for reflected light compared to transmitted light to satisfy the energy conservation. The sum of those phase shifts, or the inherent phase shift, accounts for the 180° phase difference between the probability amplitude of the $A_{tt}$ and $A_{rr}$, causing the Hong-Ou-Mandel dip. When the on-chip directional coupler has excess loss $L_{excess}$, however, the inherent phase shift will not be 180° anymore. Performing a matrix optics calculation, we have the inherent phase shift $\psi$ as $\cos(\psi) = L_{excess}^2(1+SR)^2/2SR - 1$, or $2L_{excess}^2 - 1$ for a symmetric (SR = 0-dB) directional coupler. The visibility reduction caused by the excess loss of the directional coupler can therefore be expressed as

$$1-V = \frac{L_{excess}^2(1+SR)^2}{2SR} \quad (10)$$

Here we estimate that the 0.1-dB excess loss via vertical scattering from the chip even with ideal sidewalls, or 170° internal phase shift, computed by FDTD method as noted in the earlier design section, in the balanced directional coupler will reduce the visibility by 1.5%. This excess loss will be larger when including fabrication disorder-induced losses. For unbalanced directional coupler, the internal phase shift will be further away from 180° with corresponding reductions in the visibility. Formally, the output annihilation and creation operators of a lossy directional coupler have to include Langevin noise operators to maintain the commutation relation, while at the same time inducing additional phase shifts [60].

**6. Compensation method for chip-scale two-photon interference**

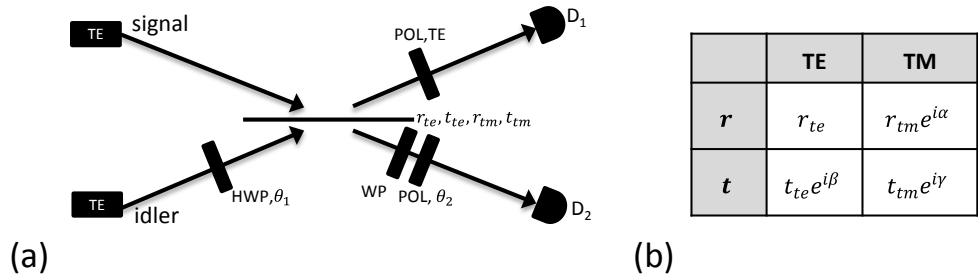

Fig. 6. (a) Compensation scenario for splitting ratio imbalanced directional coupler. The input sides are kept TE in this case. A half waveplate is inserted into the lower branch before the chip, rotating the linear polarization at $\theta_1$. One of the polarizer inserted after chip at the upper branch is rotated to TE polarization, and another at the lower branch is rotated at $\theta_2$ referencing the TE polarization. The second waveplate is inserted in one output branch in order to compensate the birefringence. (b) Nomenclature of the reflectivity and transmission for TE and TM polarization of the directional coupler. $\alpha$ denotes the birefringence at the directional coupler, $\beta$ denotes half the internal phase shift for the TE polarization, and $\gamma$ denotes half the internal phase shift for the TM polarization. In the waveguide region, the differential birefringence has a $\phi$ phase shift in the TM mode relative to the TE mode, and the differential loss from TM to TE is denoted as $l_d$. The waveplate delays the TM mode with an additional $\psi$ phase.

With the splitting ratio imbalance and excess loss of the directional coupler (due to fabrication imperfections or slight residual design), one of the co-authors has proposed an approach to compensate the imbalance and regain the indistinguishability [52], albeit for lossless and non-birefringent fiber beamsplitters. For the scalable chip implementation, a simple but lossy approach (shown in Figure 6a) would be to place half waveplates in one path before the chip and two polarizers after the chip, post-selecting the photons and removing the polarization information. When the splitting ratios are different for TE and TM modes, there exists a continuum of solutions for the angles of the polarizers to achieve the probability amplitude of $A_{rr}$ and $A_{tt}$ with the same amplitude and inverse phase. Here the distinguishability is removed as long as the following condition is met: $\tan\theta_1 \tan\theta_2 = 1 - 2\tau_{te}/\sqrt{r_{te}r_{tm}}$.

However, the birefringence of the directional coupler as well as the silicon waveguides brings additional polarization information projection to the polarizer. In more complete scenarios where loss and loss-induced internal phase shifts are considered for TE and TM modes, this distinguishability could still be compensated in our approach since only two adjustable elements, for example the angles of the polarizer and of the waveplate, are needed to recover the two probability amplitudes inverse in phase and equal in amplitude. We define the directional coupler birefringence, internal phase shifts, waveplate phase, and waveguide differential birefringence and loss in Figure 6. In this complete case, we have $A_{rr} = r_{te}(r_{te}\cos\theta_1\cos\theta_2 + l_d r_{tm} e^{i(\alpha+\varphi+\psi)}\sin\theta_1\sin\theta_2)$ and $A_{tt} = t_{te}^2 e^{i2\beta}\cos\theta_1\cos\theta_2$, where the visibility could be maximized when $A_{rr} = -A_{tt}$.

## 7. Conclusion

We have observed 1550-nm Hong-Ou-Mandel interference in silicon quantum photonic circuits, with raw quantum visibility up to 90.5% in near-symmetric directional couplers. With thermally-stabilized spectrally-bright PPKTP chip-scale waveguides as the entangled biphoton source, we examined the constituents of residual distinguishability through numerically-designed directional couplers, multiphoton pairs, polarization effects, excess loss, and imperfect phase shifts. With our sequential triggering approach for negligible coincidental dark counts, we present the theoretical analysis for multipair biphoton contribution to Hong-Ou-Mandel visibility reduction. Techniques for visibility compensation in chip-scale birefringent directional couplers in the presence of loss are described. The results presented here support the scalable realization of two-photon interaction elements on-chip, for quantum information processing and communications.

**Acknowledgments**

The authors acknowledge discussions with Fangwen Sun, Philip Battle, Tony Roberts, Xingsheng Luan, Andrzej Veitia, and Felice Gesuele. We acknowledge the scanning electron micrograph images of Figure 1 from James F. McMillan. This work is supported by the DARPA InPho program under contract number W911NF-10-1-0416.